\documentclass{elsart}

\usepackage{graphics}

\usepackage{graphicx}

\usepackage{amssymb}

\begin{document}

\begin{frontmatter}

\title{Fluxon dynamics by microwave surface resistance \\ measurements in MgB$_2$}

\author[a1]{A. Agliolo Gallitto},
\author[a1]{M. Bonura},
\author[a1]{S. Fricano}
\ead{fricano@fisica.unipa.it},
\author[a1]{M. Li Vigni},
\author[a2]{G. Giunchi}

\address[a1]{INFM and Dipartimento di Scienze Fisiche e Astronomiche, Via Archirafi 36, I-90123 Palermo (Italy)}

\address[a2]{EDISON S.p.A.-Divisione Ricerca e Sviluppo, Via U.
Bassi 2, I-20159 Milano (Italy)}

\begin{abstract}

Field-induced variations of the microwave surface resistance,
$R_s(H)$, have been investigated in high-density ceramic MgB$_2$.
At low temperatures, several peculiarities of the $R_s(H)$ curves
cannot be justified in the framework of models reported in the
literature. We suggest that they are ascribable to the
unconventional vortex structure in MgB$_2$, related to the
presence of two gaps. On the contrary, the results near $T_c$ can
be accounted for by the Coffey and Clem model, with fluxons moving
in the flux-flow regime, provided that the anisotropy of the upper
critical field is taken into due account.

\end{abstract}


\end{frontmatter}

\section{Introduction}

Since the discovery of superconductivity in the MgB$_2$ compound,
several authors have investigated the properties of such
superconductor in the presence of magnetic fields. It has been
shown that the electric, magnetic and transport properties of
MgB$_2$ are strongly affected by the magnetic field
\cite{gen1,gen2,gen3,Noi}. Such enhanced field dependence of the
properties of MgB$_2$ has been ascribed to the peculiar two-gap
structure, with a larger gap, associated with the two-dimensional
$\sigma$ band, and a smaller gap, associated with the
three-dimensional $\pi$ band, which is rapidly suppressed by the
applied magnetic field \cite{band1,band2,band3}.

Measurements of the microwave (mw) surface resistance, $R_s$,
allow investigating the processes responsible for mw energy losses
in superconductors as well as determining specific properties of
the samples. It is well known that the mw surface resistance of
superconductors in the mixed state depends on the magnetic field
through several mechanisms \cite{brandt,clem}. The different
vortex states, in the different regions of the H-T plane,
determine the temperature and field dependencies of $R_s$.
Therefore, measurements of $R_s(H, T)$ may provide important
information on the fluxon dynamics in the different regimes of
fluxon motion.

It has been shown that mw losses in MgB$_2$ superconductor are
strongly affected by the magnetic field in the whole range of
temperatures below $T_c$, even for relatively low field values
\cite{Noi,dulcic,giapan}. In particular, an unusually enhanced
field dependence of $R_s$ has been observed at $T \ll T_c$, where
pinning effects should hinder the energy losses.

In this paper, we report experimental results of the field-induced
variations of the mw surface resistance in high-density ceramic
MgB$_2$. Below $T = 0.95 T_c$, we have observed a magnetic
hysteresis in the $R_s(H)$ curves, whose peculiarities cannot
straightforwardly be justified by models reported in the
literature. On the contrary, the experimental results obtained
near $T_c$ are well accounted for in the framework of the Coffey
and Clem model \cite{clem}, with fluxons moving in the flux-flow
regime, provided that the anisotropy of the upper critical field
is taken into due account.

\section{Experimental}

The field-induced variations of $R_s$ have been studied in a bulk
ceramic MgB$_2$ sample of approximate dimensions
2$\times$3$\times$0.3 $mm^3$, with $T_c \approx$ 39 K. The sample
has been extracted from a high-density pellet (2.4 g/cm$^3$)
obtained by reactive infiltration of liquid Mg on a powdered B
preform \cite{giunchi}. After the reaction in a sealed stainless
steel container, lined with a Nb foil, a thermal treatment has
been performed for two hours at $T \approx$ 900 $^\circ$C.

The mw surface resistance has been measured using the cavity
perturbation technique. The cavity, of cylindrical shape with
golden-plated walls, is tuned in the TE$_{011}$ mode resonant at
9.6 GHz ($Q$-factor $\approx$ 40,000 at $T=4.2$ K). The sample is
put in the center of the cavity by a sapphire rod, in the region
of maximum mw magnetic field. The cavity is placed between the
poles of an electromagnet, which generates DC magnetic fields up
to $\approx$ 10 kOe. Two additional coils allow reducing to zero
the residual field and working at low magnetic fields. The quality
factor of the cavity has been measured by an hp$-8719$D Network
Analyzer. $R_s$ has been investigated as a function of the DC
magnetic field, $H_0$, in the range 0 $\div$ 10 kOe, at fixed
values of temperature. Before each measurement was performed, the
sample was zero-field cooled (ZFC) to the desired value of
temperature; the external field was then increased up to 10 kOe
and successively decreased to zero, at constant temperature. The
sample and field geometry is shown in Fig.1 a); in this geometry
the mw current induces a tilt motion of all the fluxons present in
the sample, as shown in Fig.1 b).

\begin{figure}
\centering
\includegraphics[width=70mm, height=35mm]{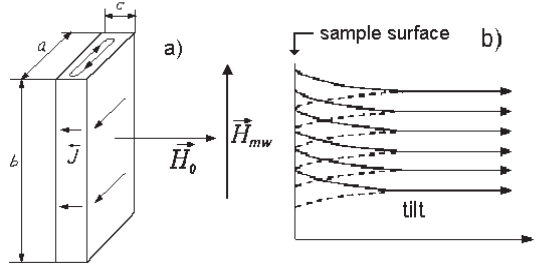}


 \caption {a) field geometry. b) fluxon motion induced by the mw
current; the arrows indicate the fluxons \cite{brandt}.}

\end{figure}

Fig.2 shows the field dependence of $R_s$, normalized to the
normal-state surface resistance, $R_n$, measured at $T$ = 40 K,
for different values of the temperature. Open and solid symbols
refer to results obtained on increasing and decreasing $H_0$,
respectively. As one can see, up to $\sim$ 4 K below $T_c$ the
$R_s(H)$ curves show a magnetic hysteresis. The inset shows the
temperature dependence of the height of the hysteresis loop at
$H_0$ = 0, $\triangle R_s(0)/R_n$, i.e. the variation of $R_s/R_n$
after a complete cycle of the DC magnetic field from 0 to 10 kOe
and back. On increasing the temperature up to $T \approx$ 34 K,
$\triangle R_s(0)/R_n$ takes a constant value; on further
increasing the temperature it quickly decreases; the hysteretic
behavior vanishes at temperatures very close to $T_c$.

Both the increasing and decreasing-field branches of the $R_s(H)$
curves show a plateau in the field ranges 0 $\div$ $H_p$ and 0
$\div$ $H^\ast$, respectively. Since the sample was ZFC, $H_p$
marks the first-penetration field. Fig.3 a) shows $H_p$ as
function of temperature; the line has been obtained by fitting the
experimental data with the law $H_p(T)=H_p(0)[1-(T/T_c)^\beta]$;
we have obtained, as best-fit parameters, $H_p(0)$ = 390 $\pm$ 20
Oe and $\beta$ = 2.4 $\pm$ 0.2. The $\beta$ value gives a
temperature dependence of $H_p$ consistent with that of
$H_{c1}(T)$. $H_p(0)$ is slightly larger than the values of
$H_{c1}(0)$ reported in the literature for MgB$_2$, suggesting
that in our sample surface-barrier effects are weak.

\begin{figure}[h]
\centering
\includegraphics[width=70mm, height=135mm]{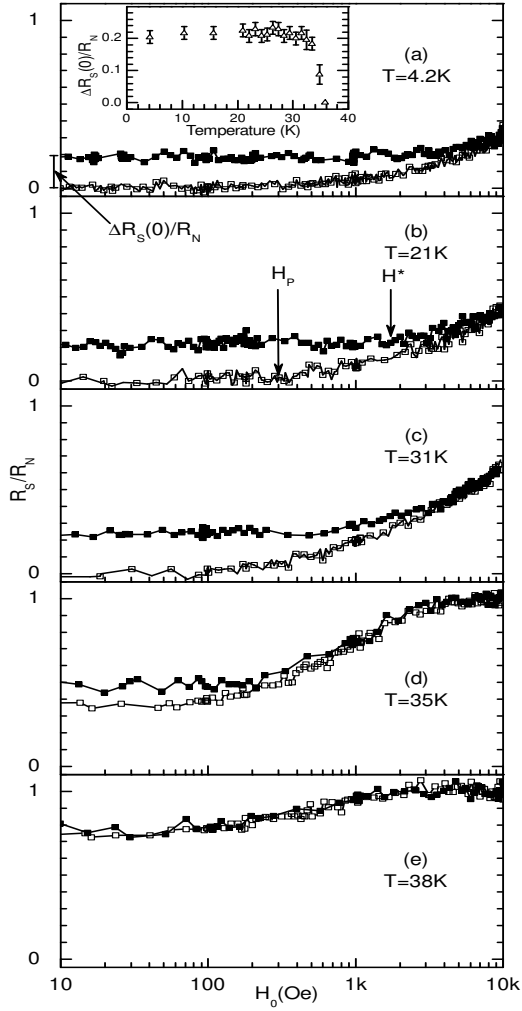}


\caption {Magnetic-field dependence of $R_s$ at different values
of temperature. Open and solid symbols refer to measurements
performed at increasing and decreasing field, respectively. The
inset shows the height of the hysteresis loop at $H_0$ = 0 as a
function of temperature.}

\end{figure}

Fig.3 b) shows the temperature dependence of $H^\ast$, though the
meaning of this characteristic parameter is not clear, one can
note that its temperature dependence is very similar to that of
$H_p$.

\begin{figure}
\centering
\includegraphics[width=65mm, height=65mm]{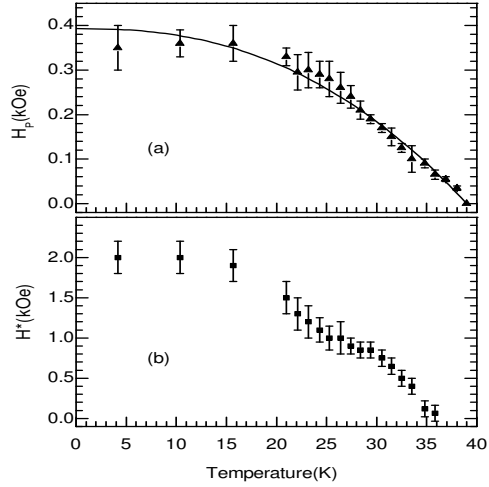}


\caption {Temperature dependence of the characteristic fields
$H_p$ and $H^\ast$, defined in Fig.2.}
\end{figure}

\section{Discussion}

Microwave losses induced by DC magnetic fields in superconductors
in the mixed state are proportional to the complex penetration
depth of the mw field, $\widetilde{\lambda}$; they are influenced
by the fluxon motion and the very presence of vortices which bring
along normal fluid in their core \cite{clem}.
$\widetilde{\lambda}$ can be expressed in terms of the
normal-fluid density, flux-flow resistivity and $\nu / \nu_0$
ratio, where $\nu$ is the working frequency and $\nu_0 = \alpha_L
/ \eta$ the depinning frequency, with $\alpha_L$ the Labush
parameter and $\eta$ the viscous-drag coefficient. For $\nu \gg
\nu_0$ the induced mw current makes fluxons moving in the
flux-flow regime. Otherwise, the fluxon motion is ruled by the
strength of the restoring-pinning forces.

Our results show that for $T <$ 35 K the $R_s(H)$ curves exhibit
an hysteretic behavior, suggesting that pinning effects are
important in this range of temperatures. On the contrary, for $T
>$ 35 K the hysteresis vanishes and fluxons should move in the
flux-flow regime. Actually, the results near $T_c$ can be
accounted for by using the Coffey and Clem theory \cite{clem},
with fluxons moving in the flux-flow regime, provided that the
anisotropy of the upper critical field is taken into due account.
Following the same procedure of Ref.\cite{Noi}, based on the
assumption that the anisotropic Ginzburg-Landau theory is valid
for MgB$_2$ in this range of temperatures, we have fitted the
experimental data by taking the anisotropy factor, $\gamma$, as
fitting parameter. The best-fit curves, along with the
experimental data, are shown in Fig.4. For all the curves, the
best-fit value of the anisotropy factor is $\gamma$ = 3, showing
that in our sample the anisotropy of $H_{c2}$ is constant in a
range of temperatures of about 3 K below $T_c$.

\begin{figure}[h]
\centering
\includegraphics[width=65mm, height=45mm]{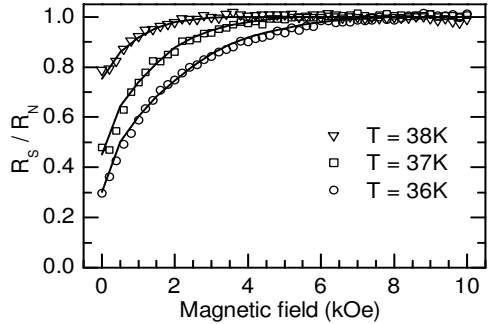}


\caption {Magnetic-field dependence of the normalized surface
resistance at temperatures close to $T_c$. Lines are the best-fit
curves obtained as described in the text.}

\end{figure}

The hysteresis of the $R_s(H)$ curves, observed for $T <$ 35 K,
should be related to the different value of the induction field at
increasing and decreasing fields. Both the normal fluid density
and the flux-flow resistivity depend on the number of fluxons
moving under the action of the mw current. In the field geometry
we used the mw current induces a tilt motion of all the fluxons
present in the sample (see Fig.1). So, the hysteresis of $R_s(H)$
should be directly related to the hysteretic behavior of the
magnetization curve due to pinning effects. However, several
anomalies can be highlighted in the $R_s(H)$ curves at low
temperatures. Firstly, from Fig.2 one can see that an unusually
enhanced field variation of $R_s$ is observed, even at the lowest
temperatures; a magnetic field of 10 kOe induces an $R_s$
variation of about 1/3 of the normal-state value. We suggest that
such enhanced field-induced energy losses are ascribable to the
effects of the two-gap structure of MgB$_2$. Indeed, it has been
shown that the structure of vortices in MgB$_2$ is characterized
by two coherence lengths, associated with each band
\cite{band2,overlap}. On increasing the applied field, the giant
cores start to overlap at field values much smaller than $H_{c2}$.
Therefore, modest fields can suppress superconductivity between
the vortices, resulting in an enhancement of the normal-fluid
density \cite{band2,band3,overlap}. Furthermore, this particular
vortex structure can modify the interaction among vortices and
pinning centers, giving rise to an unconventional shape of the
pinning potential well.

Another anomaly concerns the shape of the decreasing-field branch
of the $R_s(H)$ hysteresis loop. By supposing fluxons in the
critical state, after the external field starts decreasing the
variation of $B(H)$ should be initially slow and then faster, no
matter the field dependence of the critical current. So, a
downward curvature of the decreasing-field branch of the $R_s(H)$
curve is expected, at variance with our results. The origin of
this anomalous behavior, as well as the presence of the wide
plateau in the decreasing-field $R_s(H)$ curves, is not yet
understood and needs further investigation.

In conclusion, we have reported a detailed study of the
field-induced variations of the mw surface resistance in
high-density ceramic MgB$_2$. An unusual behavior of the $R_s(H)$
curves has been highlighted at low temperatures, which cannot be
explained in the framework of models reported in the literature.
We suggest that it is ascribable to the unconventional vortex
structure related to the presence of the two gaps. On the
contrary, the results obtained at temperatures close to $T_c$ have
been accounted for in the framework of the Coffey and Clem model,
with fluxons moving in the flux-flow regime, by taking into
account the anisotropy of upper critical field.

\end{document}